# THE COMMODIFICATION OF OPEN EDUCATIONAL RESOURCES FOR TEACHING AND LEARNING BY ACADEMICS IN AN OPEN DISTANCE E-LEARNING INSTITUTION

Lancelord Siphamandla Mncube, University of South Africa, ncubels@unisa.ac.za

Maureen Tanner, University of Cape Town, mc.tanner@uct.ac.za

Wallace Chigona, University of Cape Town, wallace.chigona@uct.ac.za

**Abstract:** The use of open educational resources (OER) is gaining momentum in higher education institutions. This study sought to establish academics' perceptions and knowledge of OER for teaching and learning in an open distance e-learning (ODeL) university. The study also sought to establish how perceptions are formed. The inductive approach followed the lens of commodification to answer the research questions. The commodification phase allowed for a better understanding of the academics' prior knowledge, informers, academics behaviour about OER, and how they perceived OER to be useful for teaching and learning. The study employed a qualitative method, with semi-structured interviews to collect data. The study found that academics with prior experience and knowledge of OER are more successful in the use of these resources for teaching, learning, and research. OER is also perceived as a useful tool to promote African knowledge, showcase the contributions of African academics, improve academic research capabilities, improve student's success rate, particularly for financially vulnerable students. Based on the acquired perceptions, the study able to propose a new guideline to formulate user perceptions. However, this can only be achieved through a solid OER policy with the support of government and tertiary institution top management. The findings may inform higher education institutions when they consider the development of OER strategies and policies, especially in response to the Covid-19 emergency online learning transition.

**Keywords:** Commodification; Open educational resources; Academics' perceptions; Teaching and learning; Open distance e-learning institution, Covid-19.

## 1. INTRODUCTION

Globally, in the diverse higher education institutions there has been an emergence of adoption and development of open educational resources (OER). Institutions comprehend these educational resources as useful and the idea of OER is sold and marketed into academics as facilitators and knowledge producers in academic spaces (Nusbaum, Cuttler & Swindell, 2019; Anderson & Cuttler, 2020). This relates to the commodification process as referred to something which converted to a commodity or a creation of a market for things so that are bought and sold (Fleissner, 2009; Pankova & Khaldeeva, 2017). The commodification process is popular in the investigation of information technology adoption (Silverstone, Hirsch & Morley, 1992), lately in the delivery of teaching and learning. Commodification has contributed to research and has contributed to the growth of social inequality (Adair, 2010). In the event of commodification, this might also create obstacles for academics as are expected to develop and adopt OER, yet it is not known if they are all comfortable with the initiative.

OER are described as open digital educational resources that are accessible from online platforms (Cobo, 2013). Also, OER are lauded for reducing the cost of teaching and learning resources in higher education institutions (Blaschke, Müskens & Zawacki-Richter, 2020). They showed to have a positive impact on cost reduction as they are free of charge, are simple to use (McGowan, 2020). These OER are created and guided by creative commons which are open licensing built within the framework of intellectual property rights of open copyright licenses. These permit to retain, reuse, revise, remix and redistribute OER (McGowan, 2020). Open educational resources are available in different formats such as open textbooks, open-source software, videos, pictures and many more (Cobo, 2013).

In many developed countries, higher education institutions are well equipped with electricity, information communication technology (ICT) infrastructure, connectivity and financial stability (McGowan, 2020). However, despite the availability of such relevant resources, the adoption of OER has not gained wider popularity across faculties and disciplines (Genc & Kocdar, 2020). This might be due to the nascent stage of OER in higher education institutions, resulting in academics not having identified appropriate strategies to adopt OER in their disciplines (McGowan, 2020). That is exacerbated in developing countries where various factors hinder the adoption and utilisation of OER by academics in higher education institutions. Barriers to OER adoption in developing countries include the poor ICT infrastructure as well as low and unaffordable internet connectivity (Hodgkinson-Williams, Arinto, Cartmill & King, 2017). Besides, there are social as well as OER-related challenges including "lack of interest, pedagogical challenges, social norms, lack of relevance, lack of institutional support, lack of capacity, lack of legal clarity, and copyright issue" (Cox & Trotter, 2017; Wolfenden & Adinolfi, 2019). These circumstances result may contribute to a low OER development and adoption rate. It is evidenced by OER portals report that the OER adoption in sub-Sahara Africa was meagre 2% (Hilton III, Bliss, Robinson & Wiley, 2013).

It is, therefore, important to explore further how academics perceive the use of OER for teaching and learning as well as their knowledge on the matter, to contribute to the commodification rate. OER are an essential initiative within the online and distance learning sphere and in other institutions. Until recently, most universities assumed a face-to-face teaching model. However, due to the Covid-19 pandemic, the tertiary education landscape has seen a rapid and unprecedented shift towards an online teaching model (World Economic Forum, 2021). Open educational resources could play an essential role in this transition. This further confirms the urgent need to understand academics' perceptions and knowledge as related to the commodification of OER to improve the effectiveness of existing teaching, learning, and research approach in an online context.

The latest open distance e-learning (ODeL) approach relies on ICT and information systems to deliver and administer teaching and learning including research. At the same time, an ODeL approach is more like a business model (University of South Africa (UNISA) Annual report, 2013). ODeL as a business model embraces only online learning provision (UNISA Annual report, 2013). An ODeL institution, therefore, expects its academics to be versed with modern ICT to provide teaching and guide students within the information systems. Teaching in an ODeL is achieved using multimedia technologies such as videos, virtual platforms, social media, video conferencing, instant communications, OER and many more ICT services. That poses challenges to ODeL academics and institutions because not all academics possess the required skills for tuition provision. Hence, the shift may pose a serious operational risk (UNISA Annual report, 2013). In an ODeL environment, everything resides in an online system and no physical human interaction occurs. Also, students are not grouped into lecture halls for lessons. Instead, they are scattered all over the world and can catch up with lessons at any given time regardless of access devices and contexts. All academics employed for teaching and learning in an ODeL institution are compelled to be innovative in the provision of subject lessons. Thus, in an evolving ICT-driven context such as ODeL, academics are expected to rely on OER instead of prescribed textbooks (de Hart, Chetty & Archer, 2015).

Given the research problems, the study sought to understand academics perceptions and knowledge of the development and adoption of OER in the context of an ODeL university which is the UNISA because in such context academics are providing teaching and learning (tuition) in an online mode and are expected to consider the variety of online teaching resources. The study posed the following key research questions:
- What are the academics' perceptions about OER for teaching and learning?
- How are the perceptions formulated?

The study employed a qualitative case study approach. The commodification process was deemed to be the most appropriate theoretical lens as discussed in the literature review. The study findings might help ODeL institution and any other higher education institution when making decisions in the strengthening or development of OER policies that will be adopted during the Covid-19 era and in the future. Besides all positive and negative factors attribute to the adoption of OER. Due to the coronavirus (Covid-19) pandemic which started from China in the city of Wuhan in 2019 (Chinazzi, Davis, Ajelli, Gioannini, Litvinova, Merler, Piontti, Mu, Rossi, Sun & Viboud, 2020). Consequently, this pandemic has claimed the lives of several people not only in China but worldwide. Based on the covid-19, various many countries initiate lockdowns trying to eliminate the spread of the virus (Bhattacharyya, Bhowmik, Mukherjee, 2020; Dhama, Sharun, Tiwari, Dadar, Malik, Singh & Chaicumpa, 2020). This lead to stop operations in many institutions such as schools, universities, travel agencies, entertainment centers, and services in the countries. Therefore, the ODeL institution might be in a better position to fight against the spread of covid-19. This can be achieved because in an ODeL institution there are fewer physical human social interactions as students and academics interact through virtual spaces of teaching and learning (Mncube, Dube & Ngulube, 2017). This influenced the investigation on how academics in an ODeL institution commodifying OER for tuition and research before and during the Covid-19 era.

## 2. LITERATURE REVIEW

The study reviewed literature aligned to the research question and the commodification process which is concerned about the user's previous phenomenon knowledge, and perceptions of IT artifacts (Berker, Hartmann, Punie & Ward, 2006). In inquiring about user's prior knowledge, it is deemed to be necessary to interrogate literature starting from the origins of OER and overview of OER. That enables the study to generalise about academics' prior knowledge. Also, that lays a foundation before starting to acquire about academics' perceptions concerning OER.

### 2.1 Commodification process

Commodification refers to how a commodity is designed and marketed to consumers (Berker, et al., 2006; Chigona, Chigona, Kayongo & Kausa, 2010). Therefore, commodification can be considered as a theoretical lens that enables the exploration of an emerging IT artifact while seeking the users' perceptions in the adoption of OER for teaching, learning, and research. According to Habib and Sønneland (2010), the commodification process "encompasses the various activities that transform new or unfamiliar commodities into objects that have the potential to raise interest in the mind of their potential users". That relates to the ODeL institution as a role player in the promotion of OER to academics, so they become aware and wildly used for tuition and research. Furthermore, it is a process through which material and symbolic artifacts are created and opened to the influence of the consumers (Chigona, et al., 2010). In the study context, OER ideas are sold to influence academicians to recognise them as an important commodity to perform their daily duties.

The commodification assisting to discover the way users experience technologies, what technologies mean to them, and how technologies play a role in their daily lives (Silverstone, Hirsch & Morley, 1992). The relevance of the commodification process might determine how academics perceive an emerging OER to be useful in their academic spaces. In the current case, commodification might provide an insight on how to promote the development and use of OER in ODeL institutions. Since many academic institutions are now obligated to make provisions for online teaching, learning and research whilst relying on ODeL modes, findings from this study are opportune. Based on this understanding, there is a link to the perceptions and knowledge of OER by the means of 'commodification'. Therefore, commodification refers to the adoption and knowledge of OER by academics where technologies initiatives, like OER, are marked with a specific function and identity within the ODeL environment.

### 2.2 The origins of OER

OER was established to address the myriad of challenges of many higher education institutions in the developing world. The common challenges these institutions were facing included access limitations, lack of adequate bandwidth, poor ICT infrastructure, inadequate local telecommunication infrastructure, and regulatory policies to govern OER as a means of access to knowledge (UNESCO. 2002). These challenges impacted student learning and success, the quality and assessment of student learning, and their affordability for higher education (León, Tejero, Dévora & Pau, 2020). The use of OER positively impacts institutional financial sustainability through increased student retention (Fine & Read, 2020).

There is confusion in the use of the terms "open electronic resources" and "open educational resources". The former is considered as library materials obtained from databases, magazines, archives, theses, conference papers, government papers, scripts, and monographs in an electronic form (Hawthorne, 2008). The term open electronic resources have existed for quite some time as it was developed in the late 1960s and was made available to the public in 1972 to enable remote online access to database services (Bates, 1998). On the other hand, the term 'open educational resources' was coined in 2002 (UNESCO. 2002). OER refers to 'internet resources', this implies an online resource that provides educational information free of charge with ease of access, due mainly to the free software and public access systems (Tucker, 2020). These terms convey the same message. Interestingly, all the content of open electronic resources informs the open educational resources. The assumption may be that the term 'Open educational resources' is more related to content and educational matters. Serious attention is needed from the higher education institutions to initiate policies related to OER, to avoid any confusion that can occur, particularly with unpacking 'open educational resources in their relevant context.

### 2.3 Overview of OER in academia

In higher education institutions academics are developing and adopting OER. Academics are expected to facilitate tuition and research and for that purpose, they may adopt and generate several teaching resources (Tsakonas & Papatheodorou, 2006). Some of the resources are OER that are produced in different formats such as but not limited to texts, images, videos, simulation, and courses (Hilton, 2016). Academics may also utilise OER that are considered as freely accessible to academic content like open access books, open sources, articles, and any subject-specific source (Hawthorne, 2008). Free access to academic content aims to enhance the use of information, electronic data management, mobile data collection, scholarly publication, and education (Tucker, 2020).

Globally, it is noted that not all academics are exposed or knowledgeable about OER (Anderson & Cuttler, 2020). The lack of knowledge might have been caused by various factors such as lack of infrastructure, inequalities in access to education (Cox & Trotter, 2017) and the fact that OER is emerging technology (Anderson & Cuttler, 2020). Furthermore, the use of OER as a possible tool for research and supervision has not yet been fully exploited to widen access to knowledge in certain aspects of research training (South African Department of Education, 2003).

### 2.4 Academics' perceptions concerning OER

Perceptions are known as cognitive and psychological processes initiated by Fiske and Taylor (1991). Lately, perceptions are considered as a process of selecting, organising, and interpreting information (Miller & Poston, 2020). In this context, the perceptions are acquired from academics concerning OER. Since the initiation of OER in 2002, there has been a shift in academics' perceptions concerning the use of OER in education. OER is increasingly perceived to be on par with traditional textbooks from a quality perspective (Blaschke, Müskens & Zawacki-Richter, 2020). OER is also perceived to be able to widen access to, reduce the costs of, and improve the quality of education (Creative commons, nd). In a recent study that investigated both traditional textbooks and OER, academics were found to have positive perceptions about OER (Maboe, 2019). Academics are positive because students who use OER can score better grades and have lower failure rates than those using traditional textbooks. Students' attitude towards the subject matter also improves when they use OER (Serrano, Dea-Ayuela, Gonzalez-Burgos, Serrano-Gil & Lalatsa, 2019).

Negative perceptions emerge when academics must search, select, edit, and apply OER as this process can be challenging (Blaschke, Müskens & Zawacki-Richter, 2020). As such, adequate guidelines to that effect may improve the academics' perceptions towards OER (Serrano, et. al., 2019). Therefore, there is a need to further explore academics' perceptions concerning OER in a developing country context is more present, given the prevailing covid-19 pandemic. As many institutions are expected to rely on an online teaching model, OER could be increasingly prioritised by academics.

## 3. THE CONTEXTUAL SETTING OF THE STUDY

The study context is the University of South Africa (Unisa). The institution consists of various campuses across nine South African provinces and beyond the country. The institution was established in 1873 (Unisa website, 2021). In the year 2000 undertook a restructuring of tertiary education; this process led to the establishment of a single distance education institution. This was approved by the minister of higher education in 2003 (South African Department of Education 2003). The institution offered distance learning through virtual platforms and little physical interaction with students and mostly relied on a blended mode of facilitation (Unisa website 2021). This became the largest open distance learning institution in Africa and the longest standing dedicated distance education university in the world (Unisa website 2021). It enrolled nearly one-third of all South African students as it relied on online and a blended mode of service delivery (Unisa website 2021). In 2013 the university began shifting from ODL to ODeL (UNISA Annual report, 2013). ODeL allows students to pursue their studies by fully relying on an online learning mode (UNISA Annual report, 2013).

The use of OER aligns with the requirements of ODeL institutions. These OER are also supported by ICT and any other information systems (Goodman, Melkers & Pallais, 2019). This is in line with ODeL institutions that rely on online information systems such as learning management systems. This means students can enroll wherever they are using online platforms for learning. Therefore, the ODeL together with OER might be the best option to increase the opportunities for people to access education (Davis & Cartwright, 2020).

Besides enabling the increase of enrolment, ODeL has also been praised for reducing enrolment costs (Blaschke, Müskens & Zawacki-Richter, 2020) and improve equity in higher education. Online and distance learning has risen in popularity in the last two decades as it is an effective approach for accommodating an increasingly diverse student population in higher education and enriching the learning environment by incorporating online teaching resources (Serrano, et al., 2019). The crisis remains because in investigations of OER development by faculties there is little knowledge produced, although studies of OER development structures have been undertaken in international settings (Hodgkinson-Williams, Arinto, Cartmill & King, 2017). The institution amended its curriculum policy to allow the usage of OER and a reduction of prescribed textbooks to support students who may not afford the books (de Hart, Chetty & Archer, 2015).

# 4. METHODOLOGY

The study opted for a case study as a research design. The study focused on a single case study which is the Unisa. This study is interpretative and employed for a qualitative approach. A qualitative inquiry was appropriate for studying commodification (Silverstone, 2005). The researchers wanted to establish academics perceptions concerning the commodification of OER. The qualitative approaches are based on the ontological assumption in which reality is understood as subjective. This means that studying perceptions and experiences that may be different for each person and change over time and context (Eriksson & Kovalainen, 2008).

The study followed a guideline in Figure 1. After inquiring and gathering findings based on the three aspects of the commodification process which are academic knowledge, informants about commodity, and user behaviour, there was a need to establish how perceptions were formed. The process started from acquiring academic knowledge (either prior or current) about the phenomenon. From there it was necessary to know how they are behaving towards the OER. Lastly, the study gathered academic behaviour toward the OER. The investigation into the three aspects helped to suggest the appropriate strategy of showing the formulation process of perceptions in an Information System organisation.

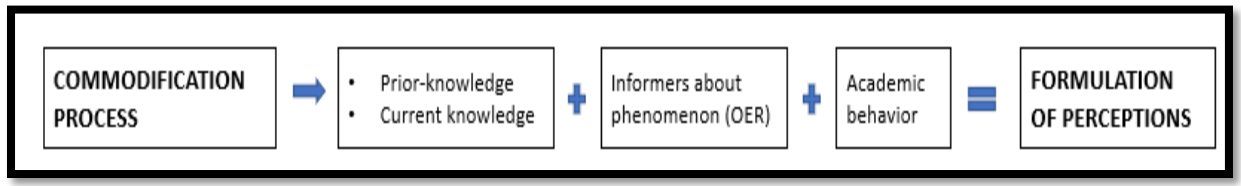

**Figure 1: Guidelines for formulation of user perceptions**

## 4.1. Sample and sampling

The study used purposive and snowball sampling. The sample was drawn from a broad heterogeneous ODeL context as consisted of eight colleges (Unisa website, 2021). Initially, the e-mails were sent to all colleges in the academics' departments to request permission to conduct semi-structured interviews. The first respondents in the respective colleges assisted to identify other relevant respondents who were involved in the adoption and development of OER. Those academics who were considered as users, adopters and developers of OER then were approached for interviews. Respondents were academics from all ranks such as junior lecturer, lecturer, senior lecturer, and professors. To qualify for participation, these academics had to be involved in teaching and learning or teaching any subject within the ODeL context. This was done because they are in a better position to commodifying OER. A total of 19 academics participated in the study.

## 4.2. Data collection and analysis

The data were collected using semi-structured interviews. This allowed the respondents to talk about and explain issues that they felt were important in their own words (Longhurst, 2010). The research instrument was based on the commodification process. The data was collected in 2019 -2020. The data collection was done in face-to-face interviews and online interviews using virtual platforms such as MS Teams. The interviews typically took place on different campuses of ODeL in the academics' offices. The interviews lasted 30 – 60 minutes. During the data collection, the data were recorded and were later transcribed into a text format. The transcribed data were coded into NVIVO for analysis. The study used thematic analysis.

## 4.3. Ethical considerations

This study is part of a PhD study that is registered at the University of Cape Town (UCT). ethical clearance for the study was obtained from the university (UCT, 2021). Further, the researchers obtained permission from the ODeL university to research within their institutions following their policy (UNISA, 2013). Permission to conduct interviews was obtained on condition that all ethical procedures be adhered to such as, anonymity including the protection of respondents, rights to participate, and their right to withdraw at any time if they felt like doing so.

## 5. EMPIRICAL FINDINGS

### 5.1. Demographic information of respondents

Table 1 presents the demographic information of the respondents.

**Table 1: Overview of demographic respondents**

| Characteristics | Respondents |
|---|---|
| **Academics gender:** | |
| Male | 14 |
| Female | 5 |
| **Academic title:** | |
| Mr | 6 |
| Mrs | 4 |
| Dr | 6 |
| Prof | 3 |
| **Qualifications:** | |
| Honours | 4 |
| Masters | 9 |
| PhD | 6 |
| **Work experience (years):** | |
| 1 – 3 | 5 |
| 4 – 5 | 5 |
| 6 – 10 | 5 |
| 11 – 15 | 1 |
| 16 – 20 | 2 |
| 21 – 25 | 0 |
| 26 – 30 | 1 |
| **Position:** | |
| Junior Lecturer | 4 |
| Lecturer | 6 |
| Senior Lecturer | 6 |
| Professor | 3 |
| **Colleges:** | |
| Accounting Sciences | 4 |
| Agriculture and environment sciences | 2 |
| Economics and management sciences | 3 |
| Education | 2 |
| Human sciences | 4 |
| Science engineering and technology | 2 |
| Graduate studies | 2 |
| **Total** | **19** |

### 5.2. *Academics' previous knowledge about open educational resources*

The majority of the respondents had previous knowledge of OER. However, the majority of the respondents did not have a clear common understanding of the OER. Some viewed OER as any other electronic or internet resource relevant to education. Some thought that there is no difference between open electronic resources and open educational resources.

> *"This is my ignorance I thought that they were open electronic resources, and I didn't know there were open educational resources"* (L5).

> *"...OER is all about having free access to academic content for academics and students can utilise in their academic space"* (SL2).

This finding concurs with literature as the OER terminology has not yet been clarified (Hawthorne, 2008; Pete, Mulder, Neto & Omollo, 2018). This might raise concerns about the term 'OER', as respondents understood OER is being relevant but were confused by what they referred to. Besides, two junior lecturers had no previous views about the OER phenomenon. This might be attributed to their work experience because many junior lecturers are new in academia.

> *"No, I experienced the OER in 2018 luckily when I arrived here in* [ODeL institution], *I was involved in a team that was responsible for developing OER so that familiarized me with the knowledge of OER"* (JL2).

### 5.3. *Sources from which academics learned about OER*

The respondents learned about OER from various sources both within the ODeL institution and other institutions. Table 2 indicates some sources of OER knowledge for academics.

**Table 2: The source of knowledge about OER**

| Sources of knowledge about OER | Some related responses |
|---|---|
| **ODeL institution** | *"I heard about OER from [ODeL institution] when the institution was advocating for OER. It was also through presentations, especially among colleagues in the health studies. They make use of OER frequently"* (L2). <br><br> *"I read from ODeL OER strategy about OER, and I don't have any problems"* (SL1) |
| **Alumni institutions** | *"When I was doing my Masters of ICT in education, exposed me further on OER and its value to education"* (L2). |
| **Open and online universities** | *".... on the internet, when I went into the internet, I saw those adverts that you can study online…without having a teacher in front of you, so that`s how I learned about them [OER]"* (SL2). |

Most respondents (World Economic Forum, 2021.) heard about the OER from the university through either chair of departments, departments, other academics, and word of mouth within their work environments. Some academics heard about it when they were redeveloping their modules and were encouraged by the institution to adopt or develop OER for their module contents. Other academics heard about OER from the university OER strategy (short guidance proposed by the institution for OER) which was approved by the institution.

Some respondents were exposed to the technology through the institutions where they previously studied. L2 learned about OER during his previous studies when he was enrolled for a qualification in the field of Education. Some academics heard about OER through internet pop-ups or wikis that advertise OER in open universities. Open universities use the system for core student activities that require the provision of OER. The respondents heard from institutions that are well established in the development of OER and that are advertised on the internet.

### 5.4. *Academics perceptions about OER for tuition and research*

Most respondents had positive perceptions about OER. Table 3 summarises some of the perceptions of academics.

**Table 3: The emerging key perceptions about OER in an ODeL**

| Emergent perceptions | Some related responses |
|---|---|
| **Effective in improving the pass rate** | *"I began using OER after realizing through a colleague that they are very useful so I had to adapt it into my portal so that I can get the very same results that my colleague got, I mean my colleague's students were passing very well through the use of OER"* (JL1). |
| **Tool for decolonising** | *"Firstly, as an activist scholar I want to change that South African academic are using African knowledge…by promoting our languages and cultures…all OER I produced are written in IsiZulu"* (SL1). |
| | *"So, for me, the potential of OER is that we can create African knowledge, African ways of doing things"* (P2). |
| **An enabler in the development of study material** | *"...when we re-developed our studying material, we were made aware that we should make use of OER and we should make use of it in our study materials because of the value it can provide in the enrichment"* (L3). <br><br> *"It is when I started using OER when I was developing online modules"* (L7). |
| **Useful in attracting students to virtual platforms** | *"It was the module media studies, I was teaching, where I used Facebook as a platform of teaching"* (L1). |
| **Useful for research and supervision** | *"Say when you are doing supervision and you get something on research proposal writing, you just copied and forward it to your students"* (Professor 1). |
| **Response to Covid-19 pandemic requirements** | *"We`ve got no choice and now that there`s Covid-19 we don't have any choice we have to use OER"* (Professor 6). |

Some respondents viewed OER as a special initiative for teaching. They considered OER as a great resource for improving the pass rate because such resources provide students with access to a broad range of knowledge. Even if students cannot afford textbooks, they could have OER as a substitute or complement to the compendium of teaching and learning resources. Some academics began to use OER as part of the open sources for teaching information systems modules. They opined that this was done to enhance access to teaching resources because of the high costs of subscribing to textbooks. Some academics perceived OER as a good initiative to promote scholarship. Besides, some academics began to develop OER because they wanted to increase and promote African knowledge which is written in indigenous languages. For example, one academic had already started to produce OER in IsiZulu (one of the widely spoken languages in South Africa).

The majority of respondents begun using OER when they were involved in the processes of developing study material for their courses. As some started using OER when they were updating and revising the content and they had no option but to use OER. As part of development, academics use OER when they were initiating the process of developing tutorial letters and developing their subject matter on module content. Tutorial letters refer to those documents prepared for each module offered in an ODeL institution for communicating with registered students on issues to do with teaching, learning, and assessment (Unisa website 2021). Some academics adopted OER because they wanted to encourage students to develop an interest in online learning. After all, some students were computer literacy and preferred virtualization of content. They opined that the decision to use OER was spurred on by the need to attract students to use the online platforms for learning that the ODeL institution subscribed to. The respondents hinted that teaching using online platforms encourages students to utilize OER. With regards to the promotion of virtual content, some of the academics have done something by embedding OER in social media and learning platforms. Some respondents adopted OER for their research because of the convenience of such resources in providing answers on methodologies underpinning scientific research processes. This indicates that OER enables a successful research process in any undertaken research. The academics encouraged the use of OER among their postgraduate students so that they could strengthen their research capacity. OER were noted as an appropriate initiative to ease the burden of supervising postgraduate students.

## 6. DISCUSSION

The study sought to find out how academics in an ODeL institution commodify OER for teaching and learning. The study noted that OER is an emerging phenomenon in an ODeL institution. OER is a new initiative and considered pivotal in providing free access to teaching and learning resources (Tucker, 2020). These OERs are capable to provide a variety of subject content for diverse disciplines through online databases. Some academics lack ICT knowledge and find it difficult to adapt and create OER for their teaching modules because they are newly employed in the virtual institution. Literature also affirms that the lack of experience and knowledge of OER negatively affected the adoption and development of OER (Goodman, Melkers & Pallais, 2019). Academics who have been employed for less than two years appear to not have more knowledge of OER in contrast to those who have been in the institution for more than five years. This might be that academics who have more than five years' experience had more OER training than those who are two years' experience. Therefore, an assertion can be made that the academics with more work experienced are successful in the adoption and development of OER for teaching, learning, and research.

Many academics in developing countries do not have the experience to develop and adopt OER (Hodgkinson-Williams, et al., 2017; Cox & Trotter, 2017; UNESCO, 2002). And yet, the recent emergence of OER is described as one of the main solutions to help institutions deliver quality learning resources for free (Anderson & Cuttler, 2020; Blaschke, Müskens & Zawacki-Richter, 2020). OER is essential in the improvement of success rate in teaching because such resources provide academics with access to a range of information (Serrano, et.al, 2019; Wiley, et al., 2017). It is, therefore, essential to devise mechanisms to inform academics from traditional face-to-face tertiary institutions in developing countries of the benefits of OER and how best to utilise these resources to rapidly improve the quality of teaching, learning and research endeavors. Without such mechanisms, the benefits of OER cannot be realised.

Also, the institution needs to be committed to driving the OER initiative with support from both government and institutional top management through an approved OER strategy. The findings confirmed that when the university is encouraging academics to create and use OER for tuition and research, this commodification of OER initiative can be successful.

The development of study material must transform since online and e-learning have gained popularity over the last two decades. This will enable the institution to operate fully online and phase out the hardcopy resources. This is seen as an effective approach for accommodating an increasingly diverse student population and enriching the learning environment through the incorporation of online teaching resources (Hilton, 2016). Hence, there is a significant need to redevelop taught modules that promote the infusion of OER into all subjects in higher education institutions. However, this can only be achieved through a clearly defined OER policy rather than an OER strategy at the institutional level. These findings give rise to the following proposition:

**Proposition 1:** *ODeL and any other higher education institutions should implement a solid OER policy with the support of government and top management to ensure the success of the OER initiatives for teaching, learning and research.*

The findings also showed that when OER becomes more prevalent in academic spaces, the higher their impact in creating a knowledge society i.e., people or individuals who have access to desired information to make informed decisions. Their prevalence might also mitigate the exclusion of students from accessing information due to high costs. For example, academics perceived OER as an enabler for the promotion of African knowledge. Such use of OER brings freedom to academics to shift their mind and be able to decolonise the education system, particularly in the South African context. It is noted that the majority of available OER were developed in the Global North (Cox & Trotter, 2017). This affirms that the Global South consumes and relies on the knowledge which is developed from the Global North (Chavez & Kovarik, 2019). The emergence of OER in the African context puts academics in a better position to showcase their academic contributions to the world by excelling in the development of African OER. This gives rise to the following proposition:

**Proposition 2:** *The use of OER is a valuable tool to promote African knowledge and content and showcase the contributions of African academics for teaching, learning and research.*

The study found that OER is perceived as an enabler in the development of free online study material. The respondents opined that some students find it expensive to purchase scholarly resources and for them, OER provided alternative support to learning. Furthermore, OER is considered effective to improve pass rate because such resources provide students with free access to a broad range of knowledge. One can, therefore, conclude that freely available OER can bring a shift in the teaching-learning space especially for financially vulnerable students who could then be afforded the possibility to freely access relevant information sources in support of their learning.

Literature also discovered that those academics who develop and employ OER contribute to students' retention and success rate (Serrano, et al. 2019; Wiley, et al., 2017). The existence of OER enables access to educational material for free in contrast to other printed materials whose prices are increasing significantly globally (Goodman, Melkers & Pallais, 2019). This finding further asserts the relevance of OER in this economic climate following the covid-19 pandemic. During the covid-19 pandemic, academic institutions and many countries of the world were forced to stop their businesses and close academic institutions (UNESCO. 2020). In that context, the OER initiative got global attention since it could allow learning to proceed without compromising on social distancing (Owolabi, 2020). This is particularly relevant to the covid-19 situation, where academics are required to design online learning material rapidly to safeguard the academic program and ensure students' success during the lockdown and emergency online mode of teaching and learning.

The findings are in line with literature as suggested that institutions and education departments should start to be innovative and recommended "including open pedagogy, open collaboration, and open assessment should be implemented to keep the learners motivated and engaged during this long period of online learning" (Mncube, Dube & Ngulube, 2017:5). Such reconstruction of education also comes with challenges related to infrastructure, pedagogy, resources, assessment, quality assurance, student support system, technology, culture, and best practices (Owolabi, 2020). This indicates that the Covid-19 pandemic forced educators, academics, United Nations, policymakers, and governments to rethink and immediately act on new educational strategies. In so doing, OER gained popularity because literature affirms that OER initiatives must be implemented during the pandemic's era and beyond (Van Allen & Katz, 2020). In this regard, research concludes with the following emerging proposition:

**Proposition 3:** *OER are perceived as a possible solution to drive social distancing and enforces e-learning for tuition and research whenever natural disasters and transmitted diseases emerged in the world.*

## 7. CONCLUSION

This study examined the perceptions of academics of OER in an ODeL university and how the perceptions are formed. The study used the lens of commodification. The study found that academics with prior experience and knowledge of OER are more successful in the use of these resources for teaching, learning, and research. Such prior knowledge can be acquired through many years of experience in the tertiary institution, online exposure, or through alumni institutions. OER is also perceived as a useful tool to promote African knowledge written in indigenous languages and that can help African academics showcase their contributions to the rest of the world. That might show that African academics are in the process of developing OER which are suitable for their contexts. Academics also perceived OER to be useful in supporting research capabilities and the improvement of pass rates. Given the fact that OER is freely available, it was found to be useful in assisting financially vulnerable students. The fact that the institution relies on its internal learning management system for OER, that become a hindering factor for people who are outside the ODeL ecosystem. The issue of openness is not yet well achieved because some of the documents that are supposed to be openly available to university portals, were hard to retrieve. Lastly, the study findings showed that OER is perceived as a role player in the provision of a safe teaching and learning approach in the Covid-19 era. It was noted to be one of the best practices in adhering to social distancing and promote e-learning.

The study concludes with two main contributions. Firstly, it enabled the identification of gaps in the literature that have led to the suggestion of three propositions which are considered as main study contributions. The second contribution is related to how the three perceptions were formulated in the commodification process. In the circles of commodification, the study concludes that the perceptions can be formed in an organisation by considering the following factors: prior-knowledge, informers about the commodity, and user (academic) behaviour towards commodity

To deviate from other processes of formulating perceptions by Fiske and Taylor (1991). The study realise that it might be relevant to any information systems studies that want to establish user perceptions about ICT-related phenomena to apply proposed guidelines in Figure 1. Besides the perceptions of academics, it was noted that the ODeL institution had an OER strategy. The findings suggest that the strategy can be strengthened by the development of a solid OER policy and with the support of government and tertiary institution management to govern OER. Therefore, the study recommends the appropriate higher education OER model and the implementation of an OER policy to guide academics about the adoption and development of OER.

## 8. ACKNOWLEDGMENT

This research is part of a PhD study which was registered at the University of Cape Town with the title "**The domestication of open educational resources by academics in an open distance e-learning university of South Africa**". This project is funded by the National Institute for the Humanities and Social Sciences (NIHSS) and collaborates annually with the South African Humanities Deans Association (SAHUDA).